\documentclass[aps,pra,showpacs,floatfix]{revtex4} 
\usepackage{graphicx}
\begin{document} 

\title{Phase diffusion pattern in quantum nondemolition 
systems} 

\author{Subhashish Banerjee}
\email{subhashishb@rri.res.in}
\affiliation{Raman Research Institute, Bangalore - 560 080, 
India} 

\author{Joyee Ghosh and R. Ghosh}
\affiliation{School of Physical Sciences, Jawaharlal Nehru 
University, New Delhi - 110 067, India} 

\date{27 March 2007} 

\begin{abstract}
We quantitatively analyze the dynamics of the quantum phase 
distribution associated with the reduced density matrix of a 
system, as the system evolves under the influence of its 
environment with an energy-preserving quantum nondemolition 
(QND) type of coupling. We take the system to be either an 
oscillator (harmonic or anharmonic) or a two-level atom (or 
equivalently, a spin-1/2 system), and model the environment as 
a bath of harmonic oscillators, initially in a general squeezed 
thermal state. The impact of the different environmental 
parameters is explicitly brought out as the system starts out 
in various initial states. The results are applicable to a 
variety of physical systems now studied experimentally with QND 
measurements. 
\end{abstract} 


\maketitle

\section{Introduction}

The theory of open quantum systems addresses the problems of 
damping and dephasing in quantum systems by its assertion that 
all real systems of interest are in fact `open' systems, each 
surrounded by its environment. Quantum optics provided one of 
the first testing grounds for the application of the formalism 
of open quantum systems \cite{wl73}. Application of open system 
ideas to other areas of physics was intensified by the works of 
Caldeira and Leggett \cite{cl83}, and Zurek \cite{wz93} among 
others. Most such studies are based on a model describing 
quantum Brownian motion of a simple harmonic oscillator in a 
harmonic oscillator environment. In such a model studied by 
Caldeira and Leggett \cite{cl83}, the coordinate of the 
particle was coupled linearly to the harmonic oscillator 
reservoir, and it was also assumed that the system and the 
environment were initially separable. The treatment of the 
quantum Brownian motion has since been generalized to the 
physically reasonable initial condition of a mixed state of the 
system and its environment by Hakim and Ambegaokar \cite{ha85}, 
Smith and Caldeira \cite{sc87}, Grabert, Schramm and Ingold 
\cite{gsi88}, and by us for the case of a system in a 
Stern-Gerlach potential \cite{sb00}, and also for the quantum 
Brownian motion with nonlinear system-environment couplings 
\cite{sb03-2}. 

The recent upsurge of interest  in the problem of open quantum systems
is  because of  the spectacular  progress in  manipulation  of quantum
states of matter (atoms, or  bosonic or fermionic gases or molecules),
encoding, transmission and processing  of quantum information, for all
of  which understanding and  control of  the environmental  impact are
essential. For  such open quantum  systems, there exists  an important
class  of  energy-preserving measurements  in  which dephasing  occurs
without damping of the system.  This may be achieved with a particular
type of  coupling between the  system and its environment,  viz., when
the  Hamiltonian $H_S$  of the  system commutes  with  the Hamiltonian
$H_{SR}$ describing  the system-reservoir interaction,  i.e., $H_{SR}$
is  a  constant  of  motion  generated  by  $H_S$  \cite{sgc96,  mp98,
gkd01}.  This  condition  describes   a  particular  type  of  quantum
nondemolition (QND) measurement scheme.

In general,  a class  of observables that  may be  measured repeatedly
with  arbitrary  precision,  with  the influence  of  the  measurement
apparatus  on the  system  being confined  strictly  to the  conjugate
observables,  is   called  QND  or   back-action  evasive  observables
\cite{bvt80,  bk92}.    Such  a  measurement   scheme  was  originally
introduced  in the  context of  the detection  of  gravitational waves
\cite{brag75,   bvk78,  un79,   ho79,  caves80,   wm94,   zu84}.   The
experimental progress  on QND measurements has been  summarized in the
review  \cite{bo96}  and the  dynamics  of  decoherence in  continuous
atom-optical QND measurements studied  in \cite{vo98}.  In addition to
its relevance in ultrasensitive  measurements, a QND scheme provides a
way  to  prepare quantum  mechanical  states  which  may otherwise  be
difficult to  create, such  as Fock states  with a specific  number of
particles.    It  has  been   shown  that   the  accuracy   of  atomic
interferometry can be improved by using QND measurements of the atomic
populations  at the  inputs to  the interferometer  \cite{kbm98}.  QND
systems  have also  been  proposed for  engineering quantum  dynamical
evolution of a system with the help of a quantum meter \cite{ca05}. We
have  recently  studied such  QND  open  system  Hamiltonians for  two
different  models of the  environment describable  as baths  of either
oscillators or spins, and  found an interesting connection between the
energy-preserving QND Hamiltonians and the phase space area-preserving
canonical transformations \cite{sb07}.

As stated above, in the  context of energy-preserving QND systems, the
only effect of the environment on  the system is dephasing and it is a
natural question to ask about  the pattern of diffusion of `phases' in
such  a situation.  Such a  question is  particularly relevant  in the
context   of  a   number  of   practical  phase   measurement  schemes
\cite{kbm98, tw00}.

What is the precise meaning of the quantum mechanical phase? 
The quantum description of phases \cite{pp98} has a long 
history \cite{pad27, sg64, cn68, pb89, ssw90}. Pegg and Barnett 
\cite{pb89}, following Dirac \cite{pad27}, carried out a polar 
decomposition of the annihilation operator and defined a 
hermitian phase operator in a finite-dimensional Hilbert space. 
In their scheme, the expectation value of a function of the 
phase operator is first carried out in a finite-dimensional 
Hilbert space, and then the dimension is taken to the limit of 
infinity. However, it is not possible to interpret this 
expectation value as that of a function of a hermitian phase 
operator in an infinite-dimensional Hilbert space \cite{ssw91, 
mh91}. To circumvent this problem, the concept of phase 
distribution for the quantum phase has been introduced 
\cite{ssw91,as92}. In this scheme, one associates a phase 
distribution to a given state such that the average of a 
function of the phase operator in the state, computed with the 
phase distribution, reproduces the results of Pegg and Barnett. 

In this paper we address the problem of quantum phase diffusion 
and study the dynamics of the quantum phase distribution 
associated with the reduced density matrix of the system for a 
number of situations of practical importance, as the system 
evolves under the influence of its environment with an 
energy-preserving QND coupling. One may take the system to be 
either an oscillator (harmonic or anharmonic) or a two-level 
atom (or equivalently, a spin-1/2 system). The phase 
distributions associated with the quantum state for the two 
cases are defined. The environment is modeled as a bath of 
harmonic oscillators, and the impact of the environmental 
parameters is quantified for different initial states of the 
system. 

The plan of the paper is as follows. In Section II, we briefly 
discuss a generic energy-preserving QND system in the context 
of open systems \cite{bg06}. The bath is taken to be initially 
in a squeezed thermal state, from which the common thermal bath 
results may be easily extracted by setting the squeezing 
parameters to zero. In Section III, we define the phase 
distribution for an oscillator system, following Agarwal {\it et 
al.} \cite{as92}. In Section IIIA in particular, we consider a 
harmonic oscillator system in QND interaction with its 
environment \cite{tw00}. We study two different initial 
conditions, of the system starting (1) in a coherent state and 
(2) in a squeezed coherent state. In Section IIIB, we study the 
case where the system is an anharmonic oscillator, which could 
arise, for example, from the interaction of a single mode of 
the quantized electromagnetic field with a Kerr medium 
\cite{gg94, vb89}. This Hamiltonian can be expressed in terms of 
the generators of the group SU(1,1). Using the positive discrete 
series representation of this group, we construct its phase 
distribution and study it for two different situations: (1) when 
the system is initially in a Kerr state, and (2) when it is 
initially in a squeezed Kerr state \cite{gg94}. In Section IV we 
consider the phase distribution for a two-level atom, extensively 
used as a model system in quantum computation \cite{wu95, ps96, 
dd95}. Following the phase distribution of angular momentum 
systems introduced by Agarwal and Singh \cite{as96}, we construct 
and study the phase distribution of the system for three 
different initial conditions of the system, starting (1) in a 
Wigner-Dicke state \cite{rd54}, which is the atomic analogue of 
the standard Fock state \cite{at72}, (2) in an atomic coherent 
state, which is the atomic analogue of the Glauber coherent state 
\cite{at72}, and (3) in an atomic squeezed state \cite{as96, 
ds94}. In Section V we present our conclusions. 

\section{Generic QND open systems} 

We consider the following Hamiltonian describing the 
interaction of a system with its environment, modeled as a 
reservoir of harmonic oscillators, via a QND type of coupling 
\cite{bg06}: 
\begin{eqnarray}
H & = & H_S + H_R + H_{SR} \nonumber\\ & = & H_S + 
\sum\limits_k \hbar \omega_k b^{\dagger}_k b_k + H_S 
\sum\limits_k g_k (b_k+b^{\dagger}_k) + H^2_S \sum\limits_k 
{g^2_k \over \hbar \omega_k}. \label{2a} 
\end{eqnarray} 
Here $H_S$, $H_R$ and $H_{SR}$ stand for the Hamiltonians of 
the system, reservoir and system-reservoir interaction, 
respectively. $H_S$ is a generic system Hamiltonian which we 
will specify in the subsequent sections to model different 
physical situations. $b^{\dagger}_k$, $b_k$ denote the creation 
and annihilation operators for the reservoir oscillator of 
frequency $\omega_k$, $g_k$ stands for the coupling constant 
(assumed real) for the interaction of the oscillator field with 
the system. The last term on the right-hand side of Eq. (1) is 
a renormalization inducing `counter term'. Since $[H_S, 
H_{SR}]=0$, the Hamiltonian (1) is of QND type. The system plus 
reservoir composite is closed obeying a unitary evolution given 
by 
\begin{equation}
\rho (t) = e^{-{i \over \hbar}Ht} \rho (0) e^{{i \over 
\hbar}Ht} , \label{2b} 
\end{equation}
where
\begin{equation}
\rho (0) = \rho^s (0) \rho_R (0),\label{2c}
\end{equation}
i.e., we assume separable initial conditions. The reservoir is 
assumed to be initially in a squeezed thermal state, i.e., it 
is a squeezed thermal bath, with an initial density matrix 
$\rho_R (0)$ given by 
\begin{equation}
\hat{\rho}_R(0) = \hat{S} (r,\Phi) \hat{\rho}_{th} 
\hat{S}^{\dagger} (r,\Phi),\label{2d} 
\end{equation}
where
\begin{equation}
\hat{\rho}_{th} = \prod_k \left[ 1 - e^{-\beta \hbar \omega_k} 
\right] e^{-\beta \hbar \omega_k \hat{b}^{\dagger}_k  
\hat{b}_k} \label{2e} 
\end{equation}
is the density matrix of the thermal bath, and
\begin{equation}
\hat{S} (r_k, \Phi_k) = \exp \left[ r_k \left( {\hat{b}^2_k 
\over 2} e^{-i2\Phi_k} - {\hat{b}^{\dagger 2}_k \over 2} 
e^{i2\Phi_k} \right) \right] \label{2f} 
\end{equation}
is the squeezing operator with $r_k$, $\Phi_k$ being the 
squeezing parameters \cite{cs85}. We are interested in the 
reduced dynamics of the `open' system of interest $S$, which is 
obtained by tracing over the bath degrees of freedom. Using 
Eqs. (\ref{2a}), (\ref{2c}) in Eq. (\ref{2b}) and tracing over 
the bath variables, we obtain the reduced density matrix for 
$S$, in the system eigenbasis, as \cite{bg06} 
\begin{eqnarray}
\rho^s_{nm} (t) & = & e^{-{i \over \hbar}(E_n-E_m)t} e^{-
i(E^2_n-E^2_m)\sum\limits_k {g^2_k \over \hbar^2\omega^2_k} 
\sin (\omega_kt)} \nonumber\\ & & \times \exp \Bigg[ - {1 \over 
2} (E_m-E_n)^2 \sum\limits_k {g^2_k \over \hbar^2 \omega^2_k} 
\coth \left( {\beta \hbar \omega_k \over 2} \right) \nonumber 
\\ & & \times \left| (e^{i\omega_k t} - 1) \cosh (r_k) + (e^{-
i\omega_kt} - 1) \sinh (r_k) e^{i2\Phi_k} \right|^2 \Bigg] 
\rho^s_{nm} (0). \label{2g} 
\end{eqnarray}
From (\ref{2g}) we obtain the master equation as
\begin{equation}
\dot{\rho}^s_{nm} (t) = \left[ -{i \over \hbar} (E_n - E_m) + i 
\dot{\eta} (t) (E^2_n - E^2_m) - (E_n - E_m)^2 \dot{\gamma} (t) 
\right] \rho^s_{nm} (t), \label{2h} 
\end{equation}
where
\begin{equation}
\eta (t) = - \sum\limits_k {g^2_k \over \hbar^2 \omega^2_k} 
\sin (\omega_k t), \label{2i} 
\end{equation}
and
\begin{equation}
\gamma (t) = {1 \over 2} \sum\limits_k {g^2_k \over \hbar^2 
\omega^2_k} \coth \left( {\beta \hbar \omega_k \over 2} \right) 
\left| (e^{i\omega_k t} - 1) \cosh (r_k) + (e^{-i\omega_k t} - 
1) \sinh (r_k) e^{i2\Phi_k} \right|^2. \label{2j} 
\end{equation}
For the case of an Ohmic bath with spectral density
\begin{equation}
I(\omega) = {\gamma_0 \over \pi} \omega e^{-\omega/\omega_c}, 
\label{2k} 
\end{equation}
where $\gamma_0$ and $\omega_c$ are bath parameters, $\eta (t)$ 
and $\gamma (t)$ can be evaluated \cite{bg06} and we quote the 
results: 
\begin{equation}
\eta (t) = -{\gamma_0 \over \pi} \tan^{-1} (\omega_c t), 
\label{2l} 
\end{equation}
and 
\begin{eqnarray}
\gamma (t) & = & {\gamma_0 \over 2\pi} \cosh (2r) \ln 
(1+\omega^2_c t^2) - {\gamma_0 \over 4\pi} \sinh (2r) \ln 
\left[ {\left( 1+4\omega^2_c(t-a)^2\right) \over \left( 1+ 
\omega^2_c (t-2a)^2 \right)^2} \right] \nonumber \\ & & - 
{\gamma_0 \over 4\pi} \sinh (2r) \ln (1+4a^2\omega^2_c) , 
\label{2m} 
\end{eqnarray}
at $T = 0$, with $t > 2a$; 
\begin{eqnarray} 
\gamma (t) & = & {\gamma_0 k_BT \over \pi \hbar \omega_c} \cosh 
(2r) \left[ 2\omega_c t \tan^{-1} (\omega_c t) + \ln \left( {1 
\over 1+\omega^2_c t^2} \right) \right] \nonumber \\ & & - 
{\gamma_0 k_BT \over 2\pi \hbar \omega_c} \sinh (2r) \Bigg[ 
4\omega_c (t-a) \tan^{-1} \left( 2\omega_c (t-a) \right) 
\nonumber \\ & & - 4\omega_c (t-2a) \tan^{-1} \left( \omega_c 
(t-2a) \right) + 4a\omega_c \tan^{-1} \left( 2a\omega_c \right) 
\nonumber \\ & & + \ln \left( {\left[ 1+\omega^2_c (t-2a)^2 
\right]^2 \over \left[ 1+4\omega^2_c (t-a)^2 \right]} \right) + 
\ln \left( {1 \over 1+4a^2\omega^2_c} \right) \Bigg] , 
\label{2n} 
\end{eqnarray} 
for high $T$, and again with $t > 2a$. Here we have taken, for 
simplicity, the squeezed bath parameters as 
\begin{eqnarray} 
\cosh \left( 2r(\omega) \right) & = & \cosh (2r),~~ \sinh 
\left( 2r (\omega) \right) = \sinh (2r), \nonumber\\ \Phi 
(\omega) & = & a\omega, \label{2o} 
\end{eqnarray} 
where $a$ is a constant depending upon the squeezed bath. We 
will make use of Eqs. (\ref{2i}), (\ref{2j}), (\ref{2l}), 
(\ref{2m}) and (\ref{2n}) in the subsequent analysis. Note that 
the results pertaining to a thermal bath can be obtained from 
the above equations by setting the squeezing parameters $r$ and 
$\Phi$ to zero. 

\section{Quantum phase distribution for an oscillator system} 

As discussed in the Introduction, it is more convenient to deal 
with the quantum phase distribution than a hermitian quantum 
phase operator. Following Agarwal {\it et al.} \cite{as92} we 
define a phase distribution ${\cal P}(\theta)$ for a given 
density operator $\rho$ as 
\begin{eqnarray}
{\cal P}(\theta) &=& {1 \over 2\pi} \langle \theta|\rho| \theta 
\rangle, ~ 0 \leq \theta \leq 2\pi, \nonumber\\ &=& {1 \over 
2\pi} \sum\limits_{m, n=0}^{\infty} \rho_{m, n} e^{ i(n-
m)\theta},  \label{3a} 
\end{eqnarray}
where the states $|\theta\rangle$ are the eigenstates of the 
Susskind-Glogower \cite{sg64} phase operator corresponding to 
eigenvalues of unit magnitude and are defined in terms of the 
number states $|n\rangle$ as 
\begin{equation} 
|\theta\rangle = \sum\limits_{n=0}^{\infty} e^{i n \theta} 
|n\rangle. \label{3b} 
\end{equation} 
The sum in Eq. (\ref{3a}) is assumed to converge. The phase 
distribution is positive definite and normalized to unity. 

\subsection{System of a harmonic oscillator} 

For the case where the system $S$ is a harmonic oscillator with 
the Hamiltonian 
\begin{equation} 
H_S = \hbar \omega \left(a^{\dag}a + {1 \over 2} \right), 
\label{3c} 
\end{equation}
the number states serve as an appropriate basis for the system 
Hamiltonian and the system energy eigenvalue in this basis is 
\begin{equation}
E_{n}= \hbar \omega \left( n + {1 \over 2}\right) . \label{3d} 
\end{equation}
Using this in Eq. (\ref{2g}) we obtain 
\begin{equation}
\rho^s_{n,m} (t)  =  e^{-i \omega(n-m)t} e^{i (\hbar \omega)^2 
(n-m)(n+m+1) \eta(t)} e^{-(\hbar \omega)^2 (n-m)^2 \gamma(t)} 
\rho^s_{n,m} (0), \label{3e} 
\end{equation}
where $\eta(t)$ and $\gamma(t)$ are as in Eqs. (\ref{2i}) and 
(\ref{2j}), respectively. 

The Hamiltonian described here has been used by Turchette {\it 
et al.} \cite{tw00} to describe an experimental study of the 
decoherence and decay of quantum states of a trapped atomic 
ion's harmonic motion interacting with an engineered 
high-temperature `phase reservoir', which is simulated by 
random variations in the trap frequency -- changing the phase 
of the ion oscillation without changing its energy, i.e., 
adiabatically modulating the trap frequency. In such a system 
it would be interesting to construct the quantum phase 
distribution associated with the reduced density matrix of the 
system and obtain the dynamics of the phase distribution as the 
system evolves under the influence of its environment. Equation 
(\ref{3e}) when substituted in Eq. (\ref{3a}) provides us with 
the phase distribution of the harmonic oscillator system 
interacting with its environment via a QND type of interaction. 

Now we obtain the phase distributions for some physically 
interesting initial conditions of our harmonic oscillator 
system $S$. 

\subsubsection{System initially in a coherent state}

The initial density matrix of the system is 
\begin{equation}
\rho^s(0) = |\alpha \rangle \langle\alpha|, \label{3f} 
\end{equation}
where 
\begin{equation}
\alpha = |\alpha| e^{i \theta_0} \label{3g} 
\end{equation}
is a coherent state \cite{sz97}. Thus the initial density 
matrix in the system basis is 
\begin{equation}
\rho^s_{n, m}(0) =  \langle n|\alpha\rangle \langle\alpha|m 
\rangle. \label{3h} 
\end{equation} 
Now making use of the expansion of the coherent state in terms 
of the number states we get 
\begin{equation}
\langle \alpha| n\rangle = {|\alpha|^n \over \sqrt{n!}} e^{-
|\alpha|^2 \over 2} e^{-i n \theta_0}. \label{3i} 
\end{equation}
It is to be noted that each of the diagonal elements of the 
above density matrix (\ref{3h}) is given by a Poisson 
distribution. Using Eq. (\ref{3i}) (and its complex conjugate) 
in Eq. (\ref{3h}), substituting it in Eq. (\ref{3e}), and then 
using Eq. (\ref{3e}) in Eq. (\ref{3a}) we obtain the phase 
distribution as 
\begin{eqnarray}
{\cal P}(\theta) & = & {1 \over 2\pi} \sum\limits_{m, 
n=0}^{\infty} {|\alpha|^{n + m} \over \sqrt{n!m!}} e^{-
|\alpha|^2} e^{-i (m - n) (\theta - \theta_0)} e^{-i \omega(m - 
n)t} \nonumber \\ & & \times e^{i (\hbar \omega)^2 (m -n)(n + m 
+ 1) \eta(t)} e^{-(\hbar \omega)^2 (n - m)^2 \gamma(t)}. 
\label{3j} 
\end{eqnarray} 

\begin{figure}
\scalebox{1.2}{\includegraphics{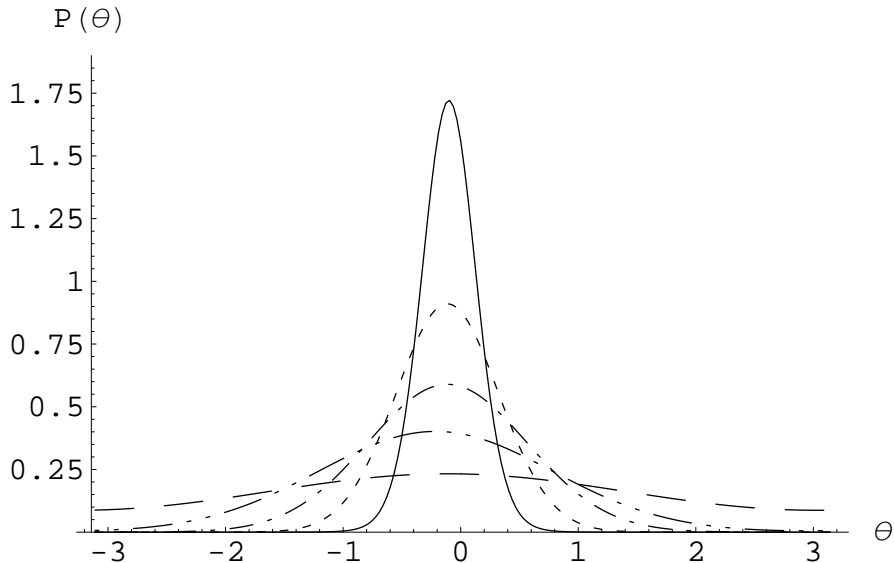}}
\caption{\scriptsize Quantum phase distribution, ${\cal 
P}(\theta)$ given by Eq. (\ref{3j}), for a harmonic oscillator 
initially in a coherent state, as a function of $\theta$ (in 
radians), for different environmental conditions and evolution 
times. The parameters have been taken as $\omega$ = 1.0, 
$\omega_c$ = 100, $|\alpha|^2$ = 5, $a$ = 0.0, $\gamma_0$ = 
0.0025. The small-dashed and the large-dashed curves are for 
temperatures $T$ (in units with $\hbar \equiv k_B \equiv 1$) = 
0 and 300, respectively, with an environmental squeezing 
parameter $r = 2$ and at an evolution time $t = 0.1$. The 
dot-dashed and the double dot-dashed curves are at evolution 
times $t$ = 0.1 and 0.2, respectively, for $T = 300$ and $r = 
1$. The continuous curve represents unitary evolution 
($\gamma_0 = 0$).} 
\end{figure} 

Figure 1 depicts the behavior of the quantum phase 
distribution, ${\cal P}(\theta)$ given by Eq. (\ref{3j}), as a 
function of $\theta$ (in radians) as it evolves under different 
environmental conditions. It can be clearly seen that in 
comparison with the unitary evolution (continuous curve), as 
the temperature $T$ increases, the phase distribution broadens 
thereby indicating increasing phase diffusion with $T$.  
The phase diffusion also increases with an increase in the 
value of the squeezing parameter $r$, defined by (\ref{2o}), as 
is evident from a comparison of the large-dashed and the 
dot-dashed curves. Also by comparing the dot-dashed and double 
dot-dashed curves indicating the same environmental conditions 
but different evolution times $t$, it can be seen that an 
increase in exposure time to the environment causes a 
corresponding increase in phase diffusion. The broadening of 
the curves in all cases takes place in such a fashion that the 
normalization of the phase distribution function is 
preserved. For all the figures in this paper, we have set 
$\omega$ = 1.0, $\omega_c$ = 100, $|\alpha|^2$ = 5, $a$ = 0.0, 
and $\theta_0$ (Eq. (\ref{3g})) = 0. 

\subsubsection{System initially in a squeezed coherent state} 

The initial density matrix of the system is 
\begin{equation}
\rho^s(0) = |\xi, \alpha \rangle \langle\alpha, \xi|, 
\label{3k} 
\end{equation}
where the squeezed coherent state is defined as \cite{sz97} 
\begin{equation} 
|\xi, \alpha \rangle = S(\xi) D(\alpha) |0\rangle. \label{3l} 
\end{equation}
Here $S$ denotes the standard squeezing operator and $D$ 
denotes the standard displacement operator \cite{sz97}. The 
initial density matrix (\ref{3k}) in the system basis is 
\begin{eqnarray}
\rho^s_{m, n}(0) & = & \langle m | \rho^s (0) |n \rangle 
\nonumber \\ & = & {e^{i{\psi \over 2}(m-n)} \over 2^{(m+n) 
\over 2} \sqrt{m!n!}} {(\tanh(r_1))^{(m+n) \over 2} \over 
\cosh(r_1)} \exp \left[-|\alpha|^2 (1 - \tanh(r_1) 
\cos(2\theta_0 - \psi)) \right] \nonumber \\ & & \times H_m 
\left[{|\alpha| e^{i(\theta_0 - {\psi \over 2})} \over 
\sqrt{\sinh(2r_1)}} \right] H^{*}_n \left[{|\alpha| 
e^{i(\theta_0 - {\psi \over 2})} \over \sqrt{\sinh(2r_1)}} 
\right] , \label{3m} 
\end{eqnarray}
where $\xi = r_1 e^{i \psi}$. Here $H_n[z]$ is a Hermite 
polynomial. Using Eq. (\ref{3m}) in Eq. (\ref{3e}) and 
substituting it in Eq. (\ref{3a}) we obtain the phase 
distribution as 
\begin{eqnarray}
{\cal P}(\theta) &=& {1 \over 2\pi} \sum\limits_{m, 
n=0}^{\infty} e^{i(n-m)\theta} {e^{i{\psi \over 2}(m-n)} \over 
2^{(m+n) \over 2} \sqrt {m!n!}} {(\tanh(r_1))^{(m+n) \over 2} 
\over \cosh(r_1)} \nonumber \\ & & \times \exp \left[-
|\alpha|^2 (1 - \tanh(r_1) \cos(2\theta_0 - \psi)) \right] 
\nonumber \\ & & \times H_m \left[{|\alpha| e^{i(\theta_0 - 
{\psi \over 2})} \over \sqrt{\sinh(2r_1)}} \right] H^{*}_n 
\left[ {|\alpha| e^{i(\theta_0 - {\psi \over 2})} \over 
\sqrt{\sinh(2r_1)}} \right] \nonumber \\ & & \times  e^{-i 
\omega(m-n)t} e^{i (\hbar \omega)^2 (m-n)(n+m+1) \eta(t)} e^{-
(\hbar \omega)^2 (n-m)^2 \gamma(t)}. \label{3n} 
\end{eqnarray} 

\begin{figure}
\scalebox{1.2}{\includegraphics{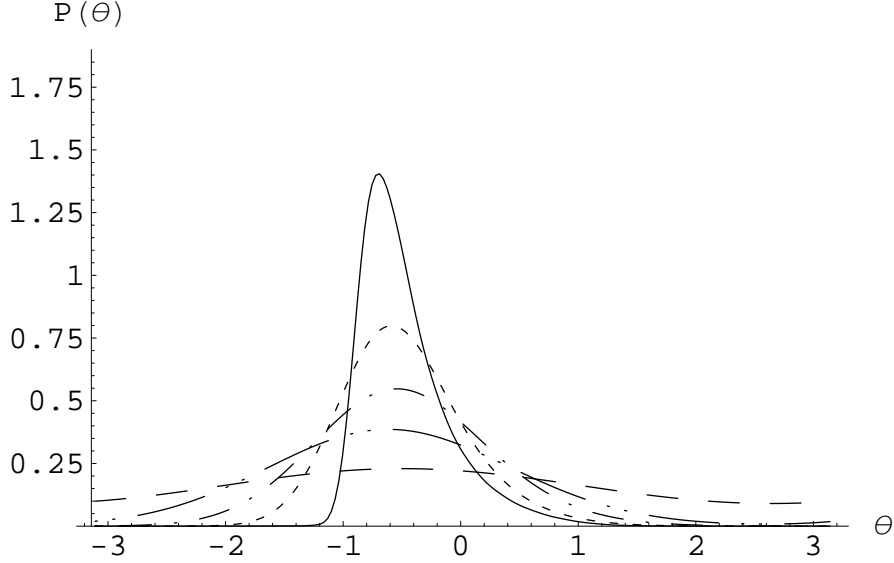}}
\caption{\scriptsize Quantum phase distribution, ${\cal 
P}(\theta)$ given by Eq. (\ref{3n}), for a harmonic oscillator 
initially in a squeezed coherent state, as a function of 
$\theta$ (in radians), for different environmental conditions 
and evolution times. The parameters have been taken as $\omega$ 
= 1.0, $|\alpha|^2$ = 5, $a$ = 0.0, $\gamma_0$ = 0.0025; $r_1$ 
=0.5, and $\psi = \pi/4$ ($r_1$ and $\psi$ are the system 
squeezing parameters (\ref{3k})). The small-dashed and the 
large-dashed curves are for temperatures $T$ (in units with 
$\hbar \equiv k_B \equiv 1$) = 0 and 300, respectively, at an 
environmental squeezing parameter (\ref{2o}) $r$ = 2 and 
evolution time $t$ = 0.1. The dot-dashed and the double 
dot-dashed curves are at evolution times $t$ = 0.1 and 0.2, 
respectively, with $T = 300$ and $r = 1$. The continuous curve 
represents unitary evolution ($\gamma_0$ = 0).} 
\end{figure} 

Figure 2 clearly indicates an increase in phase diffusion, 
corresponding to a broadening of the phase distribution curve, 
with an increase in $T$ or bath squeezing parameter $r$ or 
evolution time $t$, as was the case in Fig. 1. An interesting 
difference can be seen in the phase distribution curves 
corresponding to unitary evolution (continuous curves) in 
Figs. 1 and 2, viz., the continuous curve in Fig. 2 is more 
tilted than that in Fig. 1. This is due to the squeezing 
inherent in the initial state of the system (\ref{3k}) which is 
quantified by the parameters $r_1$ and $\psi$. 

\subsection{System of an anharmonic oscillator} 

Here we take up the case where the system is modelled as an 
anharmonic oscillator with the Hamiltonian 
\begin{equation}
H_S = \hbar \omega \left( a^{\dagger} a + {1 \over 2} \right) + 
{\hbar \lambda \over 2} (a^{\dagger})^2 a^2. \label{4a} 
\end{equation}
This has been used, for example, in studies related to a 
non-absorbing Kerr medium interacting with a single mode of the 
quantized electromagnetic field \cite{gg94,vb89}. In such a 
context $\lambda$ is related to the third-order susceptibility 
of the Kerr medium \cite{ky86}. The above Hamiltonian can be 
expressed (up to constant factors) in terms of the generators 
$K_0$, $K_+$ and $K_-$ of the $SU(1,1)$ group. These generators 
have the following bosonic representation: 
\begin{equation}
K_0 = {1 \over 4} (a^{\dagger} a + a a^{\dagger}), ~ K_+ = {1 
\over 2} (a^{\dagger})^2, ~ K_- = {1 \over 2} (a)^2. \label{4b} 
\end{equation}
In terms of these generators, Eq. (\ref{4a}) can be expressed 
as 
\begin{equation}
H_S = 2 \hbar \omega K_0 + 2 \hbar \lambda K_+ K_-. \label{4c} 
\end{equation}
We make use of the unitary irreducible representations of the 
group $SU(1,1)$ known as the positive discrete series ${\cal 
D}^{+}(k)$, where $k$ is the so called Bargmann index 
\cite{vb47}, such that the eigenvalue of the Casimir operator 
of the group is $k(k-1)$. This gives the value of $k$ to be $1 
\over 4$ or $3 \over 4$ \cite{we85, cg87}. The case of $k = {1 
\over 4}$ marks the even sector of the representation 
with the vacuum state in this representation coinciding with 
the vacuum state of the harmonic oscillator, while the case of 
$k = {3 \over 4}$ marks the odd sector of the 
representation. Thus the even and the odd sectors of the 
representation together span the number states \cite{gb00}. The 
basis for this representation obeys the following properties: 
\begin{eqnarray}
K_0|m, k\rangle &=& (m + k)|m, k\rangle, \nonumber \\ K_+|m, 
k\rangle & = & \sqrt{\left[ (m+1)(m+2k) \right]}|m+1, k\rangle, 
\nonumber \\ K_-|m, k\rangle &=& \sqrt{\left[ m(m+2k-1) 
\right]}|m-1, k\rangle, \label{4d} 
\end{eqnarray}
where $m = 0, 1, 2,... $. Using the above properties of the 
generators, the action of $H_S$ (\ref{4a}) on the basis of this 
representation is found to be 
\begin{eqnarray}
H_S |m, k\rangle &=& 2 \hbar \left[\omega(m+k) + \lambda 
m(m+2k-1)\right] |m, k\rangle \nonumber \\ & = & E_{m_k} |m, 
k\rangle. \label{4e} 
\end{eqnarray}
We use the above equation in Eq. (\ref{2g}) to obtain the 
reduced density matrix of the system $S$ in the system basis 
$|m ,k \rangle$ as 
\begin{eqnarray}
\rho^s_{m_k, n_k}(t) & = & e^{-2i(m-n)\left[ \omega + 
\lambda(m+n+2k-1)\right]t} \nonumber \\ & & \times e^{4i 
\hbar^2 (m-n)\left[\omega + \lambda(m+n+2k-1)\right] \left[ 
\omega(n+m+2k)+ \lambda (n^2 + m^2 + (2k-1)(m+n)) 
\right]\eta(t)} \nonumber \\ & & \times e^{-4 \hbar^2 (m-n)^2 
\left[ \omega + \lambda(m+n+2k-1)\right]^2 \gamma(t)} 
\rho^s_{m_k, n_k} (0). \label{4f} 
\end{eqnarray}

Now let us consider some physically interesting initial 
conditions for our anharmonic oscillator system $S$. 

\subsubsection{System initially in a Kerr state} 

An initial Kerr state $|\psi_K \rangle$ \cite{gg94} can be 
obtained as a result of an interaction of the usual coherent 
state of the electromagnetic field with a nonlinear Kerr medium 
mediated by the Hamiltonian $H_S$ given in Eq. (\ref{4a}). This 
state is defined in terms of the number states as 
\begin{equation}
|\psi_K \rangle = \sum\limits_{n} q_n |n \rangle, \label{4g} 
\end{equation} 
where
\begin{equation}
q_n = {\alpha^n \over \sqrt{n!}} e^{-|\alpha|^2 \over 2} e^{-
i \chi n(n-1)}. \label{4h} 
\end{equation}
Here $|n \rangle$ represents the usual number state and $\chi = 
{\lambda L \over 2 v}$, where $\lambda$ is as in 
Eq. (\ref{4a}), $L$ is the length of the medium and $v$ is the 
speed of light in the Kerr medium in which the interaction has 
taken place. Thus the initial system density matrix is 
\begin{eqnarray}
\rho^s_{m_k, n_k}(0) & = & \langle m, k| \psi_K \rangle \langle 
\psi_K | n, k \rangle \nonumber \\ & = & q_{2m} q^{*}_{2n} 
~~{\rm for~~~} k = {1 \over 4} \nonumber \\ & = & q_{2m+1} 
q^{*}_{2n+1} ~~{\rm for~~~} k = {3 \over 4}, \label{4i} 
\end{eqnarray}
because the state $| m, k \rangle$ with $k = {1 \over 4}$ 
represents an even number state while the state $| m, k 
\rangle$ for $k = {3 \over 4}$ represents an odd number state. 
The phase distribution is obtained by substituting 
Eq. (\ref{4i}) in Eq. (\ref{4f}) and then in Eq. (\ref{3a}), 
making use of the fact that for the positive discrete series 
representation of the group $SU(1,1)$, the even and the odd 
sectors together span the number states: 
\begin{eqnarray}
{\cal P}(\theta) & = & {1 \over 2\pi} \sum\limits_{m, 
n=0}^{\infty} q_{2m} q^{*}_{2n} e^{i2(n-m)\theta} e^{-2i(m-n) 
\left[ \omega + \lambda(m+n-{1 \over 2})\right]t} \nonumber \\ 
& & \times e^{4i \hbar^2 (m-n) \left[ \omega + \lambda(m+n-{1 
\over 2} )\right] \left[ \omega(n+m+{1 \over 2})+ \lambda(n^2 + 
m^2 - {1 \over 2} (m+n))\right]\eta(t)} \nonumber \\ & & \times 
e^{-4 \hbar^2 (m-n)^2 \left[\omega + \lambda(m+n- {1 \over 
2})\right]^2 \gamma(t)} \nonumber \\ & & + {1 \over 2\pi} 
\sum\limits_{m, n=0}^{\infty} q_{2m+1} q^{*}_{2n+1} e^{i2(n-
m)\theta} e^{-2i(m-n) \left[ \omega + \lambda(m+n+{1 \over 
2})\right] t} \nonumber \\ & & \times e^{4i \hbar^2 (m-
n)\left[\omega + \lambda(m+n+{1 \over 2} )\right] \left[ 
\omega(n+m+{3 \over 2})+ \lambda(n^2 + m^2 + {1 \over 2} 
(m+n))\right]\eta(t)} \nonumber \\ & & \times e^{-4 \hbar^2 (m-
n)^2 \left[\omega + \lambda(m+n+ {1 \over 2})\right]^2 
\gamma(t)} . \label{4j} 
\end{eqnarray}
$q_{2m}$, $q_{2m+1}$ can be obtained from Eq. (\ref{4h}). 

\begin{figure}
\scalebox{1.2}{\includegraphics{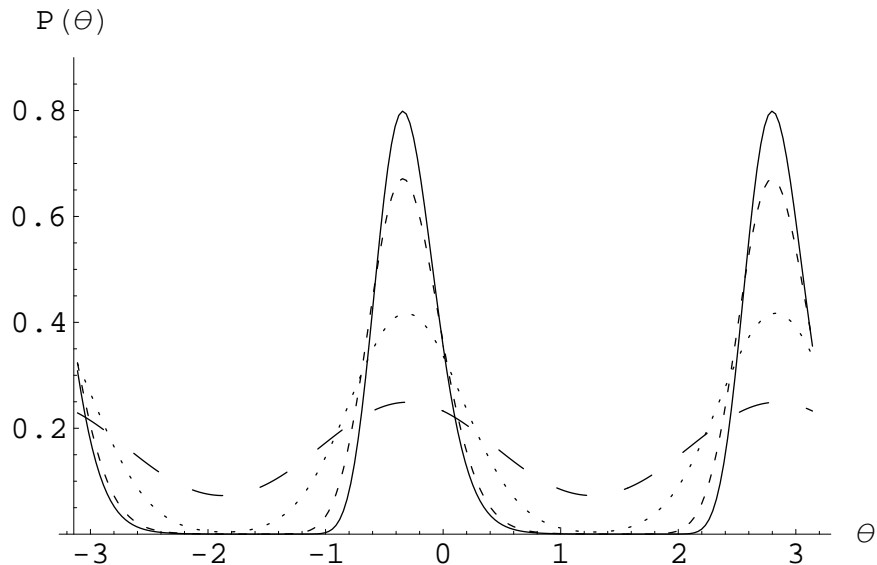}}
\caption{\scriptsize Quantum phase distribution, ${\cal 
P}(\theta)$ given by Eq. (\ref{4j}), for an anharmonic 
oscillator initially in a Kerr state, as a function of $\theta$ 
(in radians), for different environmental conditions at a fixed 
time of evolution. The parameters have been taken as $\gamma_0$ 
= 0.0025, $|\alpha|^2$ = 5, $\omega$ = 1.0, $\chi$ = $\lambda$ 
= 0.02 and evolution time $t$ = 0.1. The small-dashed and the 
large-dashed curves are for the bath squeezing parameter $r$ = 
0 and 2, respectively, at $T$ (in units with $\hbar \equiv k_B 
\equiv 1$) = 50. The dotted curve is for $T$ = 0 and $r$ = 2. 
The continuous curve represents unitary evolution ($\gamma_0 = 
0$). } 
\end{figure} 

\begin{figure}
\scalebox{1.2}{\includegraphics{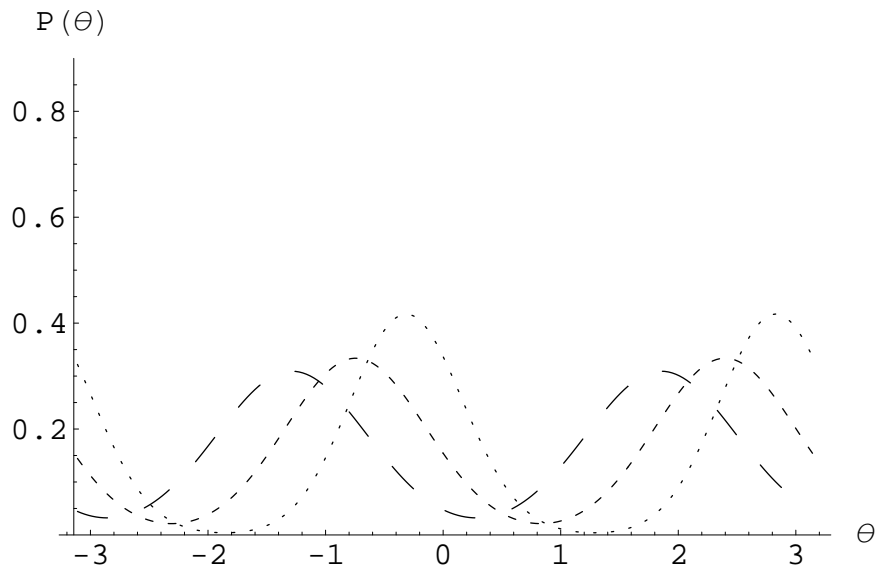}}
\caption{\scriptsize Time evolution of the quantum phase 
distribution, ${\cal P}(\theta)$ given by Eq. (\ref{4j}), for 
an anharmonic oscillator initially in a Kerr state, as a 
function of $\theta$ (in radians), for different evolution 
times under fixed environmental conditions. The parameters have 
been taken as $\gamma_0$ = 0.0025, $|\alpha|^2$ = 5, $\omega$ = 
1.0, $\chi$ = $\lambda$ = 0.02 (as in Fig. 3), $T$ = 0, and $r$ 
= 2. The dotted curve is for an evolution time $t$ = 0.1, the 
small-dashed curve is for $t$ = 0.5, and the large-dashed curve 
is for $t$ = 1.0.} 
\end{figure} 

Figures 3 and 4 represent the evolution of the quantum phase 
distribution, ${\cal P}(\theta)$ given by Eq. (\ref{4j}), as a 
function of $\theta$ for an anharmonic oscillator system 
(\ref{4a}) starting from an initial Kerr state (\ref{4g}). 
While Fig. 3 represents the evolution for a fixed evolution 
time but different environmental conditions, Fig. 4 represents 
different evolution times under the same environmental 
conditions. From Fig. 3 it is evident that increasing the 
temperature $T$ causes a broadening of the phase distributions. 
Increased phase diffusion also results from an increase in 
environmental squeezing $r$. Figure 4 clearly shows that with 
an increase in the evolution time $t$, i.e., an increase in 
exposure to the environment, the quantum phase distribution 
shifts as well as diffuses. A similar conclusion was obtained 
by Agarwal {\it et al.} \cite{as92} for an analogous situation 
studied under unitary evolution. 

\subsubsection{System initially in a squeezed Kerr state}

A squeezed Kerr state \cite{gg94} is defined in terms of the 
number states as 
\begin{equation}
|\psi_{SK} \rangle = \sum\limits_{m} s_m |m \rangle. \label{4k} 
\end{equation}
Thus the initial system density matrix in the system basis $|m, 
k \rangle$ (\ref{4e}) is 
\begin{eqnarray}
\rho^s_{m_k, n_k} (0) &=& \langle m, k| \psi_{SK} \rangle 
\langle \psi_{SK}| n, k \rangle \nonumber \\ & = & s_{2m} 
s^{*}_{2n}~~{\rm for~~~} k = {1\over 4} \nonumber \\ & = & 
s_{2m+1} s^{*}_{2n+1} ~~{\rm for~~~} k = {3\over 4}. \label{4l} 
\end{eqnarray}
Here 
\begin{equation}
s_{2m} = \sum\limits_{p} q_{2p} G _{2m2p}(z), \label{4m} 
\end{equation}
and
\begin{equation}
s_{2m+1} = \sum\limits_{p} q_{2p+1} G _{2m+12p+1}(z), 
\label{4r}
\end{equation}
with $z=r_1 e^{i\phi}$, and $G_{mp}(z) = \langle m | S(z) | p 
\rangle$, where $S(z)$ is the usual squeezing operator, is 
given by \cite{sm91} 
\begin{eqnarray}
G_{2m2p} & = & {(-1)^p \over p! m!} \left({(2p)! (2m)! \over 
\cosh(r_1)}\right)^{1 \over 2} \exp{\left(i(m - p)\phi \right)} 
\nonumber \\ & & \times \left({\tanh(r_1) \over 2} 
\right)^{(m+p)} F^2_1 \left[-p, -m; {1 \over 2}; -{1 \over 
(\sinh(r_1))^2}\right]. \label{4n} 
\end{eqnarray}
Similarly, $G_{2m+12p+1}(z)$ is given by 
\begin{eqnarray}
G_{2m+12p+1} &=& {(-1)^p \over p! m!} \left({(2p+1)! (2m+1)! 
\over \cosh^3(r_1)}\right)^{1 \over 2} \exp{\left(i(m - p)\phi 
\right)} \nonumber \\ & & \times \left({\tanh(r_1) \over 2} 
\right)^{(m+p)} F^2_1 \left[-p, -m; {3 \over 2}; -{1 \over 
(\sinh(r_1))^2}\right]. \label{4s} 
\end{eqnarray}
Here $F^2_1$ is the Gauss hypergeometric function \cite{ETBM}. 
We make use of even $p$ in Eqs. (\ref{4m}), (\ref{4n}), and odd 
$p$ in Eqs. (\ref{4r}), (\ref{4s}), because as has been pointed 
out in \cite{sm91}, $G_{mn}$ is nonzero only for $m, n$ either 
both even or both odd. Since $m$ is even in (\ref{4m}), it 
follows that $p$ too should be even, and similarily for 
(\ref{4r}) where $m$ is odd. For convenience, it is sometimes 
assumed that $\phi$ is zero and $z=r_1$ is real. Using 
Eqs. (\ref{4m}), (\ref{4r}), (\ref{4n}) and (\ref{4s}) in 
Eq. (\ref{4l}), substituting it in Eq. (\ref{4f}) and then in 
Eq. (\ref{3a}) we obtain the phase distribution function as 
\begin{eqnarray}
{\cal P}(\theta) & = & {1 \over 2\pi} \sum\limits_{m, 
n=0}^{\infty} s_{2m} s^{*}_{2n} e^{i2(n-m)\theta} e^{-2i(m-
n)\left[\omega + \lambda(m+n-{1 \over 2})\right]t} \nonumber \\ 
& & \times e^{4i \hbar^2 (m-n)\left[\omega + \lambda(m+n-{1 
\over 2} )\right] \left[\omega(n+m+{1 \over 2})+ \lambda(n^2 + 
m^2 -{1 \over 2} (m+n))\right]\eta(t)} \nonumber \\ & & \times 
e^{-4 \hbar^2 (m-n)^2 \left[\omega + \lambda(m+n-{1 \over 
2})\right]^2 \gamma(t)} \nonumber \\ & & + {1 \over 2\pi} 
\sum\limits_{m, n=0}^{\infty} s_{2m+1} s^{*}_{2n+1} e^{i2(n-
m)\theta} e^{-2i(m-n)\left[\omega + \lambda(m+n+{1 \over 
2})\right]t} \nonumber \\ & & \times e^{4i \hbar^2 (m-
n)\left[\omega + \lambda(m+n+{1 \over 2} )\right] 
\left[\omega(n+m+{3 \over 2}) + \lambda(n^2 + m^2 +{1 \over 2} 
(m+n))\right]\eta(t)} \nonumber \\ & & \times e^{-4 \hbar^2 (m-
n)^2 \left[\omega + \lambda(m+n+{1 \over 2})\right]^2 
\gamma(t)} . \label{4p} 
\end{eqnarray}
Here $s_{2m}$ is as in Eq. (\ref{4m}) and $s_{2m+1}$ is as in 
Eq. (\ref{4r}). 

\begin{figure}
\scalebox{1.2}{\includegraphics{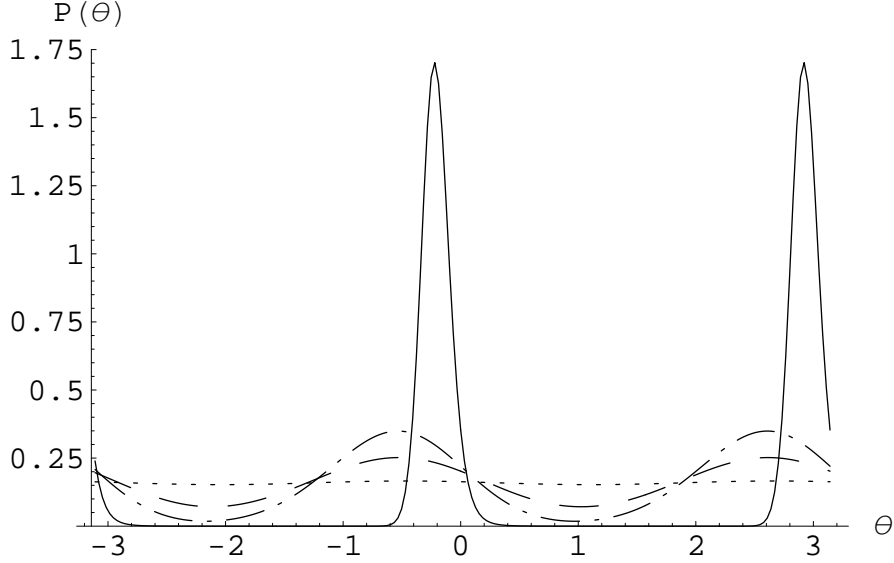}}
\caption{\scriptsize Quantum phase distribution, ${\cal 
P}(\theta)$ given by Eq. (\ref{4p}), for an anharmonic 
oscillator initially in a squeezed Kerr state, as a function of 
$\theta$ (in radians), for different environmental conditions. 
The parameters have been taken as $t$ = 0.1, $\chi = \lambda$ = 
0.02, $\gamma_0$ = 0.025, $r_1$ = 0.4 and $\phi$ = 0 ($r_1$, 
$\phi$ are the system squeezing parameters). The dot-dashed and 
the dotted curves are for $T$ (in units with $\hbar \equiv k_B 
\equiv$ 1) = 0 and 100, respectively, with the environmental 
squeezing parameter $r = 1$. The large-dashed curve is at $T$ = 
100 and $r$ = 0. The continuous curve represents the unitary 
evolution ($\gamma_0 = 0$). } 
\end{figure} 

Figure 5 depicts the evolution of the quantum phase 
distribution, ${\cal P}(\theta)$ given by Eq. (\ref{4p}), as a 
function of $\theta$ for an anharmonic oscillator system 
(\ref{4a}) starting from an initial squeezed Kerr state 
(\ref{4k}). The environmental effects are clearly depicted in 
that an increase in temperature $T$ and squeezing parameter $r$ 
results in the broadening of the phase distribution indicating 
increased phase diffusion. In the same figure, a drastic 
influence of the environment on the unitary behavior can be 
seen from the sharp fall in the amplitude of the phase 
distribution with the inclusion of environmental effects. A 
comparison between the unitary evolutions (continuous curves) 
of Figs. 3 and 5 highlights the difference in the initial 
conditions of the system depicted in these curves. The peak 
amplitude of the unitary evolution (continuous curve) is 
greater in case of a system initially in a squeezed Kerr state 
(Fig. 5) than that in a Kerr state (Fig. 3). This is indicative 
of the additional squeezing in the initial state for Fig. 5. 
The corresponding narrowing of the peaks in Fig. 5 is due to 
the fact that the phase distributions are normalized. Note that 
the multiple peaks are a common feature of the quantum phase 
distributions of the anharmonic oscillator system (\ref{4a}) as 
opposed to the single peaks of the harmonic oscillator system 
(\ref{3c}). 

\section{Quantum phase distribution of a two-level atomic 
system} 

In this section we discuss the case where our system $S$ is a 
two-level atom with a representation of the group $SU(2)$. The 
system Hamiltonian, to be substituted in Eq. (\ref{2a}), is 
\begin{equation}
H_S = {\hbar \omega \over 2} \sigma_Z, \label{5a} 
\end{equation}
where $\sigma_Z$ is the usual Pauli matrix (as has been used, 
for example, in the quantum computation models in \cite{wu95, 
ps96, dd95}). The Wigner-Dicke states \cite{rd54, jr71, at72} 
$|j, m \rangle$, which are the simultaneous eigenstates of the 
angular momentum operators $J^2$ and $J_Z$, serve as the basis 
states for $H_S$ and we have 
\begin{eqnarray}
H_S|j, m \rangle & = & \hbar \omega m |j, m \rangle \nonumber 
\\ & = & E_{j,m} |j, m \rangle. \label{5b} 
\end{eqnarray} 
Here $-j \leq m \leq j$. Using this basis and the above 
equation in Eq. (\ref{2g}) we obtain the reduced density matrix 
of the system as 
\begin{eqnarray}
\rho^s_{jm,jn}(t) &=& e^{-i \omega (m-n)t} e^{i(\hbar \omega 
)^2 (m^2 - n^2) \eta(t)} \nonumber \\ & & \times e^{-(\hbar 
\omega )^2 (m - n)^2 \gamma(t)} \rho^s_{jm,jn}(0). \label{5c} 
\end{eqnarray} 
Following Agarwal and Singh \cite{as96} we introduce the phase 
distribution ${\cal P}(\phi)$, $\phi$ being related to the 
phase of the dipole moment of the system, as 
\begin{equation} 
{\cal P}(\phi) = {2j+1 \over 4 \pi} \int_{0}^{\pi} d\theta 
\sin \theta ~ Q(\theta, \phi), \label{5d} 
\end{equation}
where ${\cal P}(\phi)> 0$ and is normalized to unity, i.e., 
$\int_{0}^{2\pi} d\phi {\cal P}(\phi) = 1$. Here $Q(\theta, 
\phi)$ is defined as 
\begin{equation}
Q(\theta, \phi) = \langle \theta, \phi|\rho^s| \theta, \phi 
\rangle, \label{5e} 
\end{equation}
where $|\theta, \phi \rangle$ are the atomic coherent states 
\cite{mr78, ap90} given by an expansion over the Wigner-Dicke 
states \cite{at72} as 
\begin{equation}
|\theta, \phi \rangle = \sum\limits_{m= -j}^j \left(\matrix{2j 
\cr j + m}\right)^{1 \over 2} (\sin(\theta / 2))^{j+m} 
(\cos(\theta / 2))^{j-m} |j, m \rangle e^{-i(j + m) \phi}. 
\label{5f} 
\end{equation} 
Using Eq. (\ref{5e}) in Eq. (\ref{5d}), with insertions of 
partitions of unity in terms of the Wigner-Dicke states, we can 
write the phase distribution function as 
\begin{eqnarray} 
{\cal P}(\phi) &=&  {2j+1 \over 4 \pi} \int_{0}^{\pi} d\theta 
\sin \theta \sum\limits_{n,m= -j}^{j} \langle \theta, \phi |j, 
n \rangle \nonumber \\ & & \times \langle j, n| \rho^s (t)| j, m 
\rangle \langle j, m| \theta, \phi \rangle. \label{5g} 
\end{eqnarray} 
We make use of Eq. (\ref{5c}) and 
\begin{equation} 
\langle j, m| \theta, \phi \rangle = \left(\matrix{2j \cr j + 
m}\right) ^{1 \over 2} (\sin(\theta / 2))^{j+m} (\cos(\theta / 
2))^{j-m} e^{-i(j + m) \phi} \label{5h} 
\end{equation} 
and its conjugate in Eq. (\ref{5g}) to obtain the required 
phase distribution for specific initial conditions of the 
system $S$. Let us now consider some physically interesting 
initial conditions for the two-level system $S$. 

\subsection{System initially in a Wigner-Dicke state}

A Wigner-Dicke state is the atomic analogue of the Fock state 
\cite{at72}. The initial density matrix of the system $S$ in 
this case is 
\begin{equation}
\rho^s(0)= |j, \tilde{m} \rangle \langle j, \tilde{m}|, 
\label{5i} 
\end{equation}
which gives
\begin{equation}
\langle j, n| \rho^s (t)| j, m \rangle = \delta_{n, \tilde{m}} 
\delta_{\tilde{m}, m}. \label{5j} 
\end{equation}
Using this, the phase distribution becomes
\begin{equation}
{\cal P}(\phi) =  {2j+1 \over 2 \pi}  \left(\matrix{2j \cr j + 
\tilde{m}}\right) B\left[j + \tilde{m} + 1, j - \tilde{m} + 1 
\right]. \label{5k} 
\end{equation}
Here $B$ stands for the Beta function. It is evident from 
Eq. (\ref{5k}) that the phase distribution for the atomic 
system starting in a Wigner-Dicke state is uniform and is 
independent of any bath dynamics. Here, since we have only one 
two-level system in Eq. (\ref{5a}), $j = {1 \over 2}$ and 
${\cal P}(\phi)$ can be seen to go over to ${1 \over 2 \pi}$, 
i.e., a uniform distribution. 

\subsection{System initially in an atomic coherent state}

An atomic coherent state is the atomic analogue of the Glauber 
coherent state \cite{at72}. The initial density matrix of the 
system $S$ in this case is 
\begin{equation}
\rho^s(0) = |\alpha, \beta \rangle \langle \alpha, \beta| 
\label{5l} 
\end{equation}
yielding the following matrix element in the $| j, m \rangle$ 
basis:
\begin{eqnarray}
\langle j, n| \rho^s(t)|j, m \rangle & = & e^{-i \omega (n-m)t} 
e^{i(\hbar \omega )^2 (n^2 - m^2) \eta(t)} \nonumber \\ & & 
\times e^{-(\hbar \omega )^2 (n - m)^2 \gamma(t)} \langle j, 
n|\alpha, \beta \rangle \langle \alpha, \beta| j, m \rangle. 
\label{5m} 
\end{eqnarray}
Using Eqs. (\ref{5m}), (\ref{5h}) in Eq. (\ref{5g}) we obtain 
the phase distribution as 
\begin{eqnarray}
{\cal P}(\phi) & = & {2j+1 \over 4 \pi} \int_{0}^{\pi} d\theta 
\sin \theta \sum\limits_{n,m= -j}^{j}  \left(\matrix{2j \cr j + 
n}\right) \left(\matrix{2j \cr j + m}\right) \nonumber \\ & & 
\times (\sin(\theta / 2))^{2j+n+m} (\cos(\theta / 2)) ^{2j-n-m} 
e^{-i(n-m)\beta} \nonumber \\ & & \times (\sin(\alpha / 
2))^{2j+n+m} (\cos(\alpha / 2)) ^{2j-n-m} \nonumber \\ & & 
\times e^{-i \omega (n-m)t} e^{i(\hbar \omega )^2 (n^2 - m^2) 
\eta(t)} \nonumber \\ & & \times e^{-(\hbar \omega )^2 (n - 
m)^2 \gamma(t)} e^{i(n-m) \phi}. \label{5n} 
\end{eqnarray} 
In the above equation, the $\theta$ integral can be carried out 
to yield 
\begin{eqnarray}
{\cal P}(\phi) & = & {2j+1 \over 2 \pi} \sum\limits_{n,m= -
j}^{j}  \left(\matrix{2j \cr j + n}\right) \left(\matrix{2j \cr 
j + m}\right) {\Gamma(j + {1 \over 2}(n + m)+ 1) \Gamma(j - {1 
\over 2}(n + m) + 1) \over \Gamma(2j + 2)} \nonumber \\ & & 
\times e^{-i(n-m)\beta} (\sin(\alpha / 2))^{2j+n+m} 
(\cos(\alpha / 2))^{2j-n-m} \nonumber \\ & & \times e^{-i\omega 
(n-m)t} e^{i(\hbar \omega )^2 (n^2 - m^2) \eta(t)} \nonumber \\ 
& & \times e^{-(\hbar \omega )^2 (n - m)^2 \gamma(t)} e^{i(n-m) 
\phi}. \label{5o} 
\end{eqnarray}
Here $\Gamma$ is the standard Gamma function. Since $H_S$ given 
by Eq. (\ref{5a}) represents a single two-level atom, $j= {1 
\over 2}$. Eq. (\ref{5o}) is thus considerably simplified and 
we obtain the phase distribution as 
\begin{equation}
{\cal P}(\phi) = {1 \over 2 \pi}\left[1 + {\pi \over 4} 
\sin \alpha \cos(\beta + \omega t - \phi) e^{- (\hbar 
\omega)^2 \gamma(t)}\right]. \label{5p} 
\end{equation}
It can be easily checked that this ${\cal P}(\phi)$ is 
normalized to unity. As can be seen from Eq. (\ref{5p}), only 
$\gamma(t)$ plays a role in carrying the effect of the 
environment on the phase distribution. For a generic QND open 
quantum system described by (\ref{2a}), it can be shown that 
$\dot{\gamma}(t)$ is the decoherence causing term \cite{bg06}. 
Thus Eq. (\ref{5p}) is a simple and neat formula clearly 
illustrating the effect of the environment on phase diffusion. 
By making use of $\gamma(t)$ from Eqs. (\ref{2m}) and 
(\ref{2n}) for $T = 0$ and for high $T$, respectively, we find 
that the second term on the right-hand side of Eq. (\ref{5p}) 
has a power-law decay at zero $T$ and an exponential decay at 
high $T$, and eventually the phase distribution tends to the 
uniform limit of $\frac{1}{2\pi}$. Thus the effect of the 
environment stays for a longer time at zero $T$ as compared to 
that at high $T$. 

\begin{figure}\scalebox{1.2}
{\includegraphics{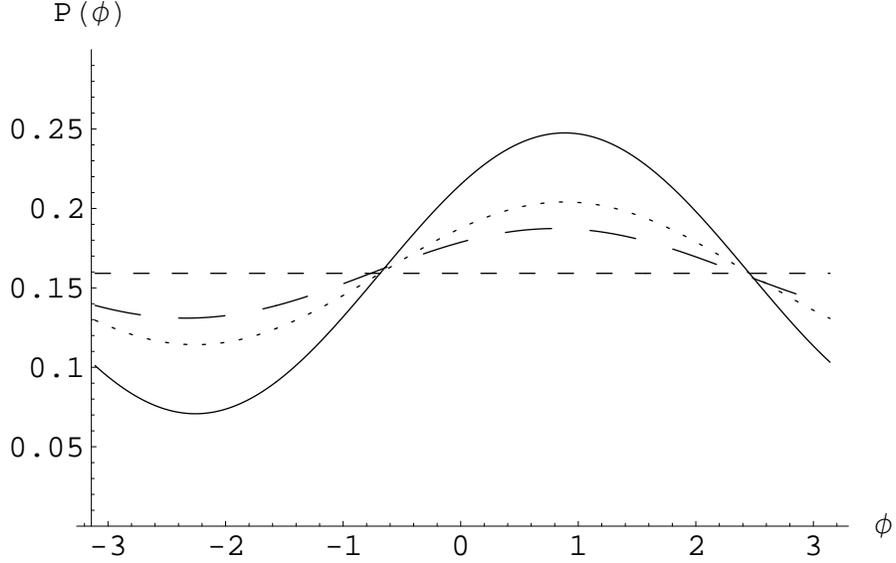}}
\caption{\scriptsize Quantum phase distribution, ${\cal 
P}(\phi)$ given by Eq. (\ref{5p}), for a two-level atom 
initially in an atomic coherent state, as a function of $\phi$ 
(in radians), for different environmental conditions and 
evolution times. The parameters have been taken as $\alpha = 
\beta = \pi/4$ [Eq. (\ref{5l})], and $\gamma_0$ = 0.025. The 
continuous curve and the dotted curve are for the bath 
squeezing parameter $r$ = 0 and 2, respectively, at a 
temperature $T$ = 0 and an evolution time $t$ = 0.1. The 
small-dashed and the large-dashed curves correspond to 
evolution times $t$ = 0.1 and 0.02, respectively, at $T$ (in 
units with $\hbar \equiv k_B \equiv$ 1) = 300 and $r$ = 2.} 
\end{figure} 

Figure 6 depicts the evolution of the quantum phase 
distribution, ${\cal P}(\phi)$ given by Eq. (\ref{5p}), as a 
function of $\phi$ (in radians) for different environmental 
conditions. It is clearly seen that increasing the temperature 
$T$, the bath squeezing parameter $r$, and the environment 
exposure time $t$ cause a broadening of the phase distribution 
curve, indicating an increase of phase diffusion. The 
broadening of the curves preserves the normalization 
of the phase distribution. 

\subsection{System initially in an atomic squeezed state}

An atomic squeezed state \cite{as76, mr78, ds94, ap90} is 
expressed in terms of the Wigner-Dicke states as 
\begin{equation}
|\zeta, p \rangle = A_p \exp(\Theta \hat{J}_Z) \exp(-i {\pi 
\over 2} \hat{J}_Y)|j, p \rangle, \label{5q} 
\end{equation} 
where
\begin{equation}
e^{2 \Theta} = \tanh(2 |\zeta|), \label{5r}
\end{equation}
with $\zeta$ indicating the initial squeezing of the system. 
The initial density matrix of the system $S$ in this case is 
\begin{equation}
\rho^s(0)= |\zeta, p \rangle \langle \zeta, p|. \label{5s} 
\end{equation}
Using Eq. (\ref{5g}) along with the expressions
\begin{eqnarray}
\langle j, n| \rho^s(t)| j, m \rangle & = & e^{-i \omega (n-
m)t} e^{i(\hbar \omega )^2 (n^2 - m^2) \eta(t)} \nonumber \\ & 
& \times e^{-(\hbar \omega )^2 (n - m)^2 \gamma(t)} \langle j, 
n|\zeta, p \rangle \langle \zeta, p| j, m \rangle, \label{5t} 
\end{eqnarray} 
and
\begin{equation}
\langle j, n|\zeta, p \rangle = A_p e^{n \Theta} d^j_{np} \left( 
{\pi \over 2} \right) , \label{5u} 
\end{equation}
\cite{mr78}, where $ d^j_{np}({\pi \over 2})$ is the standard 
Wigner symbol for the rotation operator \cite{vkam}: 
\begin{equation}
d^j_{np} \left( {\pi \over 2} \right) = \langle j, n| e^{-i{\pi 
\over 2} J_Y}| j, p \rangle, \label{5v} 
\end{equation}
and
\begin{equation} 
|A_p|^2 = \left(\sum\limits_{r}{(-1)^r (2j - r)! (\cosh \Theta 
)^{2j - 2r} \over r! (j + p - r)! (j - p -r)!}\right)^{-1}, 
\label{5w} 
\end{equation}
we obtain the phase distribution function as 
\begin{eqnarray}
{\cal P}(\phi) & = & {2j+1 \over 4 \pi} |A_p|^2 \int_{0}^{\pi} 
d\theta \sin \theta \sum\limits_{n,m= -j}^{j}  \left(\matrix{2j 
\cr j + n}\right)^{1 \over 2} \left(\matrix{2j \cr j + 
m}\right)^{1 \over 2} \nonumber \\ & & \times (\sin(\theta / 
2))^{2j+n+m} (\cos(\theta / 2))^{2j-n-m} e^{i(n-m)\phi} 
\nonumber \\ & & \times e^{-i \omega (n-m)t} e^{i(\hbar \omega 
)^2 (n^2 - m^2) \eta(t)} \nonumber \\ & & \times e^{-(\hbar 
\omega )^2 (n - m)^2 \gamma(t)} e^{(n+m) \Theta} d^j_{np} \left( 
{\pi \over 2} \right) d^{*j}_{mp} \left( {\pi \over 2} \right). 
\label{5x} 
\end{eqnarray}
In Eq. (\ref{5x}) the $\theta$ integral can be carried out to 
yield 
\begin{eqnarray}
{\cal P}(\phi) & = & {2j+1 \over 2 \pi} |A_p|^2 
\sum\limits_{n,m= -j}^{j}  \left(\matrix{2j \cr j + 
n}\right)^{1 \over 2} \left(\matrix{2j \cr j + m}\right)^{1 
\over 2} {\Gamma(j + {1 \over 2}(n + m)+ 1) \Gamma(j - {1 \over 
2}(n + m) + 1) \over \Gamma(2j + 2)} \nonumber \\ & & \times 
e^{i(n-m)\phi} e^{-i \omega (n-m)t} e^{i(\hbar \omega )^2 (n^2 
- m^2) \eta(t)} \nonumber \\ & & \times e^{-(\hbar \omega )^2 
(n - m)^2 \gamma(t)} e^{(n+m) \Theta} d^j_{np} \left( {\pi 
\over 2} \right) d^{*j}_{mp} \left( {\pi \over 2} \right) . 
\label{5y} 
\end{eqnarray} 
As discussed in subsection IVB above, for a single two-level 
system, $j = {1 \over 2}$. We take up two cases for the two 
values of $p$ appearing in Eq. (\ref{5q}): $p = -{1 \over 2}$ 
called the south pole of the phase space of the two-level 
system, and $p = {1 \over 2}$ called the north pole of the 
phase space \cite{ds94}. 

a. \underline{South pole} $(p = -{1 \over 2})$: \\ The phase 
distribution in Eq. (\ref{5y}) is considerably simplified to 
give 
\begin{equation} 
{\cal P}(\phi) = {1 \over 2 \pi} \left[ 1 - {\pi \over 4 
\cosh \Theta } \cos(\phi - \omega t ) e^{-(\hbar \omega )^2 
\gamma(t)} \right]. \label{5z} 
\end{equation} 

\begin{figure}
\scalebox{1.2}{\includegraphics{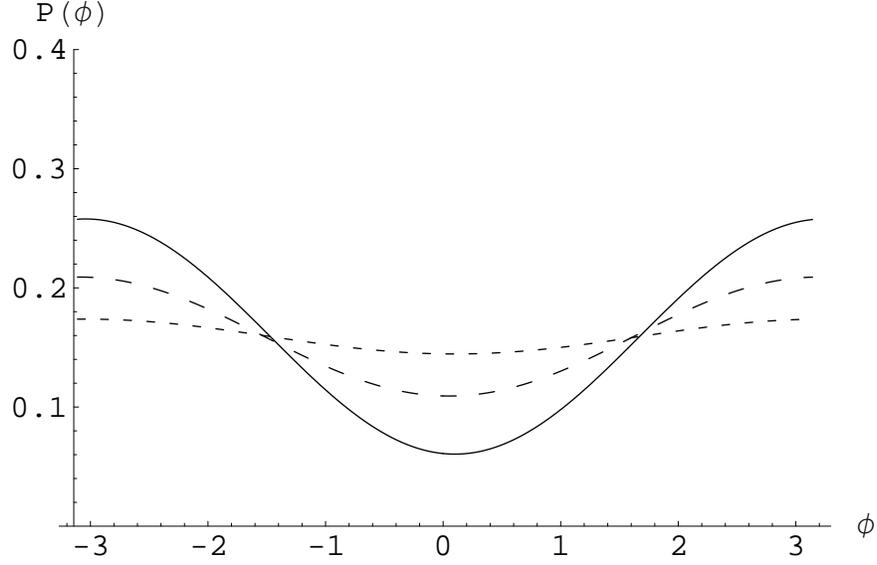}}
\caption{\scriptsize Quantum phase distribution, ${\cal 
P}(\phi)$ given by Eq. (\ref{5z}), at the south pole of the 
phase space for a 
two-level atom initially in an atomic squeezed state, as a 
function of $\phi$ (in radians), for different environmental 
conditions. The system squeezing parameter (\ref{5r}) $\Theta$ 
is taken as = - 0.5494, the bath squeezing parameter $r$ = 1, 
and $\gamma_0$ = 0.025. The continuous curve corresponds to a 
temperature $T$ = 0 and an evolution time $t$ = 0.1, while the 
small-dashed and large-dashed curves correspond to evolution 
times $t$ = 0.1 and 0.05, respectively, at $T$ (in units with 
$\hbar \equiv k_B \equiv$ 1) = 300.} 
\end{figure} 

b. \underline{North pole} $(p = {1 \over 2})$:\\ Eq. (\ref{5y}) 
is simplified to give
\begin{equation}
{\cal P}(\phi) = {1 \over 2 \pi} \left[ 1 + {\pi \over 4 
\cosh \Theta } \cos(\phi - \omega t ) e^{-(\hbar \omega )^2 
\gamma(t)} \right].  \label{72} 
\end{equation}

\begin{figure}
\scalebox{1.2}{\includegraphics{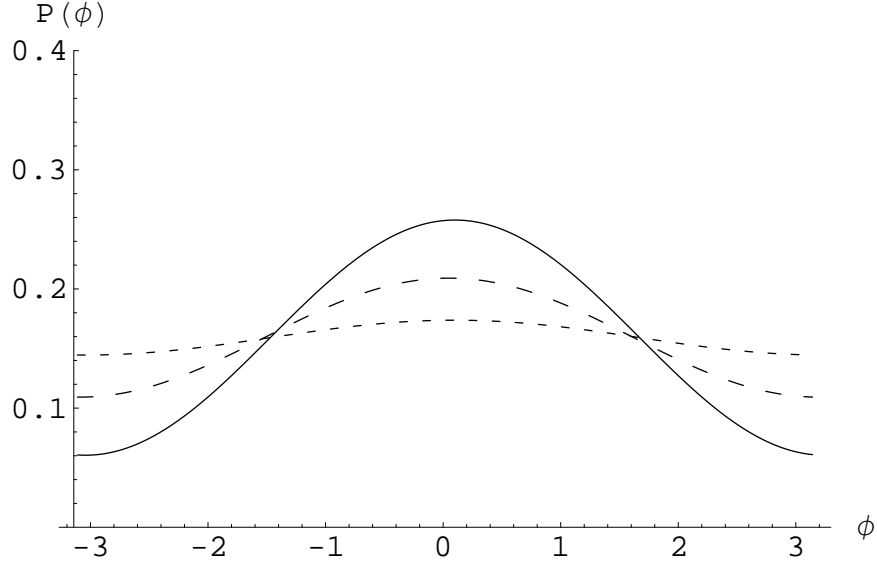}}
\caption{\scriptsize Quantum phase distribution, ${\cal 
P}(\phi)$ given by Eq. (\ref{72}), at the north pole of the 
phase space for a 
two-level atom initially in an atomic squeezed state, as a 
function of $\phi$ (in radians), for different environmental 
conditions. The system squeezing parameter (\ref{5r}) $\Theta$ 
has been taken as = - 0.5494, the bath squeezing parameter $r$ 
= 1, and $\gamma_0$ = 0.025. The continuous curve corresponds 
to a temperature $T = 0$ and an evolution time $t = 0.1$, while 
the small-dashed and large-dashed curves correspond to 
evolution times $t$ = 0.1 and 0.05, respectively, at $T$ (in 
units with $\hbar \equiv k_B \equiv$ 1) = 300.} 
\end{figure} 

$A_p$ is defined by Eqs. (\ref{5q}) and (\ref{5u}), and is 
usually fixed by normalization as in the above equations, where 
$|A_p|^2 = (\cosh \Theta)^{-1}$ which is equal to 1 for 
$\Theta$ = 0 \cite{mr78}, implying an infinite initial 
squeezing $\zeta$ of the system (\ref{5r}). 
The expression in Eq. (\ref{5z}) for the south pole in the 
phase space and that in Eq. (\ref{72}) for the north pole in 
the phase space differ from each other by a sign in the second 
part of the expressions. The contrast between 
them can be seen clearly from Figs. 7 and 8. As seen from the 
figures, with the increase in temperature $T$ or the exposure 
time to the environment $t$, the phase distribution curves 
flatten out indicating increased phase diffusion. These curves 
bring out another notable feature, viz., with the increase in 
bath exposure time $t$ or temperature $T$, the effect of 
squeezing, indicated by the parameter $r$, is washed out. This 
behavior is analogous to the effect of squeezing in oscillator 
systems, where with the increase in bath exposure time $t$ and 
temperature $T$, the non-stationary effects introduced by the 
squeezed bath are washed out \cite{bk05}. 

Eqs. (\ref{5z}) and (\ref{72}) are easily seen to be normalized 
to unity. As in subsection IVB above, the effect of the 
environment shows up in the above equations only in the function 
$\gamma(t)$, responsible for decoherence. As was the case in the 
previous subsection IVB, from the forms of the 
function $\gamma(t)$, for an Ohmic bath, given by 
Eqs. (\ref{2m}) and (\ref{2n}) for $T = 0$ and for high $T$, 
respectively, we find that the second term on the right-hand 
side of Eqs. (\ref{5z}) and (\ref{72}) has a power-law decay 
at zero $T$ and an exponential decay at high $T$, and 
eventually the phase distribution tends to the uniform limit of 
$\frac{1}{2\pi}$ seen in the case in IVA. Thus the effect of 
the environment stays for a longer time at zero $T$ as compared 
to high $T$, and eventually the distribution tends to the same 
uniform value irrespective of the initial state being a 
coherent or a squeezed state. As pointed out in \cite{as96, 
ds94, ap90}, the state $|\zeta, p \rangle$ in (\ref{5q}) has an 
inherent squeezing which is represented by $\zeta$ in 
(\ref{5r}) and the environmental squeezing $r$ is encapsulated 
in the function $\gamma(t)$ given by (\ref{2j}). Thus the 
results (\ref{5z}) and (\ref{72}) bring out the relative 
importance of the different squeezing sources and hence are 
applicable in the context of the experiment of Kuzmich {\it et 
al.} \cite{kbm98} concerning the role of the environment on the 
atomic quantum nondemolition measurements and squeezing. 

\section{Conclusions}

In this paper we have analyzed the quantum phase distribution 
of a number of physically interesting systems interacting with 
their environment via a QND type of coupling. We have taken our 
system to be either an oscillator (harmonic or anharmonic) or a 
two-level atom (or equivalently, a spin-1/2 system), and 
modeled the environment as a bath of harmonic oscillators, 
initially in a general squeezed thermal state, from which the 
common thermal bath results may be easily extracted by setting 
the squeezing parameters to zero. We have explicitly evaluated 
the phase distribution and worked out the effects of different 
environmental parameters on the dynamics of the system starting 
with various initial states. 

In particular, for a harmonic oscillator system in QND 
interaction with its environment (Section IIIA), we have 
considered two different initial conditions of the system, 
starting (1) in a coherent state, and (2) in a squeezed 
coherent state. The phase distribution corresponding to the 
unitary evolution in the second case is more tilted than that 
in the first case, which is a signature of the squeezing 
inherent in the initial state of the system. We have next taken 
an anharmonic oscillator (Section IIIB), which could arise, for 
example, from the interaction of a single mode of the quantized 
electromagnetic field with a Kerr medium, and constructed its 
phase distribution, again for different initial conditions: (1) 
the system starting in a Kerr state, and (2) the system 
starting in a squeezed Kerr state. With an increase in the 
evolution time $t$, indicating an increase in exposure to the 
environment, the quantum phase distribution in the first case 
shifts as well as diffuses. 

We have then studied the phase distribution for a discrete 
two-level atom (Section IV), for different initial conditions 
of the system, starting (1) in a Wigner-Dicke state, which is 
the atomic analogue of the standard Fock state, (2) in an 
atomic coherent state, which is the atomic analogue of the 
Glauber coherent state, and (3) in an atomic squeezed state. In 
the first test case, the phase distribution is uniform and is 
independent of any bath dynamics. In the other two cases, it is 
seen that the effect of the environment stays for a longer time 
at zero temperature than that at high temperature, and 
eventually the distribution tends to the uniform value of the 
first case, irrespective of the initial state of the system. 

In all the cases considered, a broadening of the phase 
distribution curve, indicating an increase in phase diffusion, 
results with an increase in the bath temperature $T$ or bath 
squeezing parameter $r$ or evolution time $t$. The broadening 
of the curves, of course, preserves the normalization of the 
phase distribution. Even though each system considered is an 
`open' system, we could make use of the underlying group 
symmetries of the system Hamiltonians, because of the QND 
nature of the system-environment coupling. Our quantitative 
results are of potential use in the analysis of a broad class 
of relevant experimental situations dealing with quantum 
nondemolition measurements and squeezing. 

\begin{acknowledgments}

SB would like to acknowledge R. Srikanth for useful 
discussions. The work of JG is supported by the Council of 
Scientific and Industrial Research, India. The School of 
Physical Sciences, Jawaharlal Nehru University, is supported by 
the University Grants Commission, India, under a Departmental 
Research Support scheme. 

\end{acknowledgments}


\begin{thebibliography}{99}

\bibitem{wl73} W. H. Louisell, {\it Quantum Statistical 
Properties of Radiation} (John Wiley and Sons, 1973). 

\bibitem{cl83} A. O. Caldeira and A. J. Leggett, Physica A {\bf 
121}, 587 (1983). 

\bibitem{wz93} W. H. Zurek, Phys. Today {\bf 44}, 36 (1991); 
Prog. Theor. Phys. {\bf 87}, 281 (1993). 

\bibitem{ha85} V. Hakim and V. Ambegaokar, Phys. Rev. A {\bf 
32}, 423 (1985). 

\bibitem{sc87} C. M. Smith and A. O. Caldeira, Phys. Rev. A 
{\bf 36}, 3509 (1987); {\it ibid} {\bf 41}, 3103 (1990). 

\bibitem{gsi88} H. Grabert, P. Schramm and G. L. Ingold, 
Phys. Rep. {\bf 168}, 115 (1988). 

\bibitem{sb00} S. Banerjee and R. Ghosh, Phys. Rev. A {\bf 62}, 
042105 (2000). 

\bibitem{sb03-2} S. Banerjee and R. Ghosh, Phys. Rev. E {\bf 
67}, 056120 (2003). 

\bibitem{sgc96} J. Shao, M-L. Ge and H. Cheng, Phys. Rev. E 
{\bf 53}, 1243 (1996). 

\bibitem{mp98} D. Mozyrsky and V. Privman, Journal of Stat. Phys.
{\bf 91}, 787 (1998). 

\bibitem{gkd01} G. Gangopadhyay, M. S. Kumar and S. Dattagupta, 
J. Phys. A: Math. Gen. {\bf 34}, 5485 (2001). 

\bibitem{bvt80} V. B. Braginsky, Y. I. Vorontsov and K. S. Thorne, 
Science {\bf 209}, 547 (1980).

\bibitem{bk92} V. B. Braginsky and F. Ya. Khalili, in 
{\it Quantum Measurements}, edited by K. S. Thorne 
(Cambridge University Press, Cambridge, 1992).

\bibitem{brag75} V. B. Braginsky and Yu. I. Vorontsov,
Usp. Fiz. Nauk {\bf 114}, 41 (1974)  
[Sov. Phys. Usp. {\bf 17}, 644 (1975)];
V. B. Braginsky, Yu. I. Vorontsov, and 
V. D. Krivchenkov, Zh. Eksp. Teor. Fiz. {\bf 68}, 55 (1975) 
[Sov. Phys. JETP {\bf 41}, 28 (1975)].

\bibitem{bvk78} V. B. Braginsky, Yu. I. Vorontsov and F. Ya. Khalili,
Pis'ma Zh. Eksp. Teor. Fiz. {\bf 27}, 296 (1978)  
[Sov. Phys. JETP Lett.. {\bf 27}, 276 (1978)].

\bibitem{un79} W. G. Unruh, Phys. Rev. D {\bf 19}, 
2888 (1979).

\bibitem{ho79} J. N. Hollenhorst, Phys. Rev. D {\bf 19}, 
1669 (1979).

\bibitem{caves80} C. M. Caves, K. S. Thorne, R. W. P. Drever, 
V. D. Sandberg, and M. Zimmerman, Rev. Mod. Phys. {\bf 52}, 341 
(1980). 

\bibitem{wm94} D. F. Walls and G. J. Milburn, {\it Quantum 
Optics} (Springer, Berlin, 1994). 

\bibitem{zu84} W. H. Zurek, in {\it The Wave-Particle Dualism},  
edited by S. Diner, D. Fargue, G. Lochak and F. Selleri 
(D. Reidel Publishing Company, Dordrecht, 1984). 


\bibitem{bo96} M. F. Bocko and R. Onofrio, Rev. Mod. Phys. {\bf 
68}, 755 (1996). 

\bibitem{vo98} R. Onofrio and L. Viola, Phys. Rev. A {\bf 58}, 
69 (1998). 

\bibitem{kbm98} A. Kuzmich, N. P. Bigelow and L. Mandel, 
Europhys. Lett. {\bf 42}, 481 (1998). 

\bibitem{ca05} J. Clausen, J. Salo, V. M. Akulin and 
S. Stenholm, Phys. Rev. A {\bf 72}, 062104 (2005). 

\bibitem{sb07} S. Banerjee and R. Ghosh, J. Phys. A: 
Math. Theo. {\bf 40}, 1273 (2007). 

\bibitem{tw00} Q. A. Turchette, C. J. Myatt, B. E. King, 
C. A. Sackett, D. Kielpinski, W. M. Itano, C. Monroe and 
D. J. Wineland, Phys. Rev. A {\bf 62}, 053807 (2000).  

\bibitem{pp98} V. Perinova, A. Luks and J. Perina, {\it Phase 
in Optics} (World Scientific, Singapore, 1998). 

\bibitem{pad27} P. A. M. Dirac, Proc. R. Soc. Lond. A {\bf 
114}, 243 (1927). 

\bibitem{sg64} L. Susskind and J. Glogower, Physics {\bf 1}, 49 
(1964). 

\bibitem{cn68} P. Carruthers and M. M. Nieto, 
Rev. Mod. Phys. {\bf 40}, 411 (1968). 

\bibitem{pb89} D. T. Pegg and S. M. Barnett, J. Mod. Opt. {\bf 
36}, 7 (1989); Phys. Rev. A {\bf 39}, 1665 (1989). 

\bibitem{ssw90} J. H. Shapiro, S. R. Shepard and N. C. Wong, 
Phys. Rev. Lett. {\bf 62}, 2377 (1989). 

\bibitem{ssw91} J. H. Shapiro and S. R. Shepard, Phys. Rev. A 
{\bf 43}, 3795 (1991). 

\bibitem{mh91} M. J. W. Hall, Quantum Opt. {\bf 3}, 7 (1991). 

\bibitem{as92} G. S. Agarwal, S. Chaturvedi, K. Tara and 
V. Srinivasan, Phys. Rev. A {\bf 45}, 4904 (1992). 

\bibitem{bg06} S. Banerjee and R. Ghosh, eprint quant-ph/0703054. 

\bibitem{gg94} C. C. Gerry and R. Grobe, Phys. Rev. A {\bf 49}, 
2033 (1994). 

\bibitem{vb89} V. Buzek, Phys. Rev. A {\bf 39}, 5432 (1989). 

\bibitem{wu95} W. G. Unruh, Phys. Rev. A {\bf 51}, 992 (1995). 

\bibitem{ps96} G. M. Palma, K-A. Suominen and A. K. Ekert, 
Proc. R. Soc. Lond. A {\bf 452}, 567 (1996). 

\bibitem{dd95} D. P. DiVincenzo, Phys. Rev. A {\bf 51}, 1015 
(1995). 

\bibitem{as96} G. S. Agarwal and R. P. Singh, Phys. Lett. A 
{\bf 217}, 215 (1996). 

\bibitem{rd54} R. H. Dicke, Phys. Rev. {\bf 93}, 99 (1954). 

\bibitem{at72} F. T. Arecchi, E. Courtens, R. Gilmore and 
H. Thomas, Phys. Rev. A {\bf 6}, 2211 (1972). 

\bibitem{ds94} J. P. Dowling, G. S. Agarwal and W. P. Schleich, 
Phys. Rev. A {\bf 49}, 4101 (1994). 

\bibitem{cs85} C. M. Caves and B. L. Schumaker, Phys. Rev. A 
{\bf 31}, 3068 (1985); B. L. Schumaker and C. M. Caves, Phys. 
Rev. A {\bf 31}, 3093 (1985). 

\bibitem{sz97} M. O. Scully and M. S. Zubairy, {\it Quantum 
Optics} (Cambridge University Press, Cambridge, 1997). 

\bibitem{ky86} M. Kitagawa and Y. Yamamoto, Phys. Rev. A {\bf 
34}, 3974 (1986). 

\bibitem{vb47} V. Bargmann, Ann. Math. {\bf 48}, 568 (1947). 

\bibitem{we85} K. Wodkiewicz and J. H. Eberly, 
J. Opt. Soc. Am. B {\bf 2}, 458 (1985). 

\bibitem{cg87} C. C. Gerry, Phys. Rev. A {\bf 35}, 2146 (1987). 

\bibitem{gb00} C. C. Gerry and A. Benmoussa, Phys. Rev. A {\bf 
62}, 033812 (2000); H. Ui, Progress of Theoretical Physics {\bf 
44}, 703 (1970). 

\bibitem{sm91} M. V. Satyanarayana, Phys. Rev. D {\bf 32}, 400 
(1985); P. Marian, Phys. Rev. A {\bf 44}, 3325 (1991). 

\bibitem{ETBM} A. Erdelyi, W. Magnus, F. Oberhettinger and 
F. G. Tricomi, {\it Higher Transcendental Functions}, Vol. I 
(McGraw-Hill, New York, 1953). 

\bibitem{jr71} J. M. Radcliffe, J. Phys. A: Gen. Phys. {\bf 4}, 
313 (1971). 

\bibitem{mr78} M. A. Rashid, J. Math. Phys. {\bf 19}, 1391 
(1978). 

\bibitem{ap90} G. S. Agarwal and R. R. Puri, Phys. Rev. A {\bf 
41}, 3782 (1990). 

\bibitem{as76} C. Aragone, E. Chalbaud and S. Salamo, 
J. Math. Phys. {\bf 17}, 1963 (1976). 

\bibitem{vkam} D. A. Varshalovich, A. N. Moskalev and 
V. K. Khersonskii, {\it Quantum Theory of Angular Momentum} 
(World Scientific, Singapore, 1988). 

\bibitem{bk05} S. Banerjee and J. Kupsch, J. Phys. A: 
Math. Gen. {\bf 38}, 5237 (2005). 

\end{thebibliography}
\end{document}